# ELASTIC PROPERTIES, YIELD SURFACE AND FLOW RULE OF NANOPOWDER COMPACTS

G.Sh. Boltachev[*], E.A. Chingina, N.B. Volkov, K.E. Lukyashin

Institute of Electrophysics, Ural Branch of RAS, Amundsen str. 106, Ekaterinburg, Russia

[*] e-mail: grey@iep.uran.ru

**Abstract.** Different compaction processes of the nanosized granular system, which is a prototype of an alumina nanopowder, are studied by the granular dynamics method. For all processes: compaction curves "density vs. pressure" of the powder compact are calculated, the elastic and the plastic parts are extracted from the total deformation, the body elastic moduli are determined within the isotropic solid approximation. The inadequacy of the isotropy approximation is established. The nanopowder yield surface is constructed in the space of stress tensor invariants. The inapplicability of the traditional associated flow rule for description of oxide nanopowders compaction processes is revealed. An alternative flow rule is suggested.

## 1. Introduction.

In the production of new materials with unique properties, oxide ceramic materials based on compounds such as aluminum oxide [1-8], yttrium oxide [7, 9, 10], etc. have attractive prospects. Aluminum oxide, in particular, has a high thermal conductivity and transparency, which makes it a promising candidate as a working medium for solid-state lasers [4], high strength characteristics, chemical and heat resistance, which causes its demand as a structural material [1, 2]. Increasing the strength properties, as well as improving transparency, requires reducing the average grain size in the manufactured ceramics up to values of the order of 10 nm [3, 6]. In this regard, great efforts have recently been directed to the development of nanotechnology, and in particular, the production of nanostructured ceramics by powder metallurgy methods [6-15]. One of the frequently used stages of these methods is cold pressing of nanoscale powders. Unlike micron-sized and larger powders, nanopowders have a number of unexpected properties [11, 16-18] that affect their pressing and subsequent sintering. First of all, this is a pronounced size effect: the smaller the particle size of the powder, the more difficult it is to press. Achieving the desired densities of the oxide nanopowder at the stage of cold pressing sometimes requires pressures of several GPa [11, 15-18]. In addition, as has been recently discovered [16-18], nanopowders of oxide materials are weakly sensitive to the compression scheme: the differences in density in the processes of uniform compression and uniaxial compaction do not exceed 1%. Thus, nanoscale powders are fundamentally different in their mechanical properties from conventional powder bodies.

   The rapid development of experimental techniques and further success in the production of nanostructured oxide ceramics require a corresponding development of theoretical ideas about the mechanical properties of the nanopowder compact. In the space of invariants of the stress tensor, the yielding surface of nanopowders has a convex shape of the elliptical type [12-19], which suggests the use of modified versions of the plasticity theory as a continuum approach for describing the nanopowder properties [12-15, 19, 20]. Of course, a number of formulations and terminology of the theory become quite arbitrary: in particular, the plasticity of the powder body is not connected with the deformation of individual particles but with processes of cross-slip and the repackaging. The features of the nanopowder body require a serious revision of the main formulations of the theory and verification of its results regarding the properties of the described object. A full-scale experiment cannot provide comprehensive information about the characteristics of the powder system and changes in its properties in the compaction processes. Much more detailed information can be obtained in the framework of the microscopic examination implemented in this study, i.e., the nanopowders simulation by the method of granular dynamics [16-18, 21-25].

The object of research is a monodisperse (particle diameter $d_g$ = 10 nm) model system corresponding to an aluminum oxide nanopowder with a strong tendency to agglomeration. Real powders of this type are produced in the IEP (UB of RAS, Ekaterinburg) by methods of electric explosion of conductors [26] and laser evaporation of targets [9, 27]. Individual particles are characterized by a spherical shape and high strength characteristics. The sphericity of particles and their high strength, non-susceptibility to plastic crumpling, make the method of granular dynamics particularly promising and adequate tool for theoretical analysis.

**2. Numerical simulation technique**

The model cell has the shape of a rectangular parallelepiped with sizes $x_{cell}$, $y_{cell}$, and $z_{cell}$. To generate initial backfills, the algorithm described in Ref. [16] is used, which allows creating isotropic and homogeneous structures in the form of a connected 3D periodic cluster. The number of particles $N_p$ = 4000, the initial density $\rho_0$ = 0.24. The density $\rho$ is the relative volume of the solid phase, i.e., $\rho = (\pi/6) N_p d_g^3 / V_{cell}$ where $d_g$ is the particles diameter and $V_{cell}$ is the model cell volume. Periodic boundary conditions are used on all sides of the cell. The system is deformed by simultaneously changing the sizes of the model cell and proportionally rescaling the coordinates of all particles. After each act of deformation, a new equilibrium position of the particles is determined. This procedure corresponds to the effect on the powder under quasi-static conditions. The tensor of total deformations of the model system in Cartesian coordinates is diagonal. The increment of its components at each step of deformation are related to each other by the relations: $\Delta\varepsilon_{xx} = \kappa_x \Delta\varepsilon_{zz}$, $\Delta\varepsilon_{yy} = \kappa_y \Delta\varepsilon_{zz}$. The vertical axis $Oz$ always corresponds to the maximum compression. The step of deformation along this axis for all processes is set to $\Delta\varepsilon_{zz} = \Delta z_{cell} / z_{cell} = -0.0005$, and the differences between the processes, i.e., the specifics of compaction, are determined by the values of the coefficients $\kappa_x$ and $\kappa_y$.

The stress tensor $\sigma_{ij}$ averaged over the model cell is calculated using the well-known Lava formula [21-23, 28]

$$\sigma_{ij} = \frac{-1}{V_{cell}} \sum_{k<l} f_i^{(k,l)} r_j^{(k,l)}, \qquad (1)$$

where summation is performed for all pairs of interacting particles $k$ and $l$; $f^{(k,l)}$ is the total force acting on the particle $k$ from the particle $l$; $r^{(k,l)}$ is the vector connecting the centers of the particles. As a rule, it is assumed [12-15] that the stress state of a plastically deformable body is sufficiently characterized by the first two invariants of the stress tensor, or uniquely associated with these invariants by the average (hydrostatic) stress $p$ and the intensity $\tau$ of the stress deviator ($\tau_{ij} = \sigma_{ij} - p\delta_{ij}$, where $\delta_{ij}$ is the unit tensor):

$$p = \frac{-1}{3} \text{Sp}(\sigma_{ij}), \qquad \tau^2 = \sum_{i,j=1}^{3} \tau_{ij} \tau_{ji}. \qquad (2)$$

In the following we will also introduce the axial pressures: $p_x = -\sigma_{xx}$, $p_y = -\sigma_{yy}$, and $p_z = -\sigma_{zz}$. The force characteristics of interparticle interactions are described by the relations [16-18]:

$$f_a(r) = \frac{\pi^2}{3} \frac{(nd_0^3)\varepsilon d_g^6}{(r+\alpha d_0)^3 \left[(r+\alpha d_0)^2 - d_g^2\right]^2}, \qquad (3)$$

$$\frac{f_e(r)}{Ed_g^2} = \frac{(h/d_g)^{3/2}}{3(1-v)^2} - \frac{\pi}{4}\frac{(1-v)}{(1-2v)(1+v)}\left[\frac{h}{d_g} + \ln\left(1-\frac{h}{d_g}\right)\right], \quad h = d_g - r \qquad (4)$$

$$f_t(\delta) = \min\left\{\frac{4Ea\delta}{(2-v)(1+v)}; \ \mu f_e; \ \pi a^2\sigma_b\right\}, \qquad a = \sqrt{hd_g}/2 \qquad (5)$$

$$M_p(\theta_p) = \min\left\{\frac{8Ea^3\theta_p}{3(1+v)}; \ \mu M(a); \ \frac{\pi}{2}a^3\sigma_b\right\}, \qquad M(a) = -2\pi\int_0^a \sigma_n(r)r^2 dr, \qquad (6)$$

$$M_r(\theta_r) = \min\left\{\frac{4}{3}\frac{Ea^3\theta_r}{1-v^2}; \ \frac{1}{3}af_e\right\}, \qquad (7)$$

Here: the modified Hamaker formula (3) determines the dispersion attraction force $f_a$ [29,30]; the modified Hertz's law (4) determines the force $f_e$ of elastic repulsion of particles [31]; the linearized Cataneo-Mindlin's law (5) determines the tangential interaction of pressed particles ("friction" forces) [32-34]; the linearized Jager's law (6) (or Reisner-Sagosi's law [35,36]) determines the moment $M_p$ of surface forces arising during mutual rotation of pressed particles around the contact axis at an angle $\theta_p$; the Lurie's law (7) determines moment $M_r$ surface forces that occur when the contact axis is bent at an angle $\theta_r$ (with the appearance of a strong bond between the particles; see [37], p. 272, Eq. (4.5)). In the presented equations: $r$ is the distance between the centers of the interacting particles, $\varepsilon$ and $d_0$ are the energy and the size parameters of intermolecular forces; $\alpha$ is the factor that defines the minimum gap between the contacting particles and sets a maximum adhesion force ($f_{a,\max} = f_a(d_g)$); $E$ and $v$ are Young's modulus and Poisson's ratio of the particles; $\delta$ is the tangential displacement of the contact spot; $a$ is the contact area radius; $\mu$ is the friction coefficient; $\sigma_b$ is the critical shear stress, which characterizes the shear strength of the material; $\sigma_n$ is the normal stresses on the contact area.

The appearance/destruction of a strong bond between the particles is described using a parameter $\Delta r_{ch}$ that characterizes the necessary compression of the particles [16]. It is assumed that reducing the distance $r$ between the centers of the particles to the value $r_{\min} \le d_g - \Delta r_{ch}$ initiates the formation of a strong bond. After the formation of a strong bond between the particles, further compression (with a decrease in $r$) continues to correspond to the elastic interaction (4), but when stretched (an increase in $r$) we have a linear relationship of the force $f_e$ and the distance $r$ up to the value $r' = r_{\min} + \Delta r_{ch}$. At $r > r'$ a partial contact destruction is introduced, which is described by increasing the $r_{\min}$ parameter, so that the difference $r - r_{\min}$ remains equal to its maximum value $\Delta r_{ch}$. Complete destruction of the contact between the particles occurs when stretched to the value of $r = d_g$. With the appearance of a strong bond between the particles, the restrictions in the ratios (5) and (6) associated with the friction coefficient ($\mu$) are removed.

Aluminum oxide in the α-phase is assumed as the particle material: $E = 382$ GPa, $v = 0.25$, $nd_0^3 = \sqrt{2}$, $\alpha d_0 = 0.1$ nm; $\varepsilon = 1224 k_B$, $\sigma_b = 0.02E$ [16]. The particle diameter $d_g = 10$ nm, the interparticle friction coefficient $\mu = 0.1$, the parameter $\Delta r_{ch} = 0.01 d_g$. Thus, the simulated system is close in its parameters to the model system II of Ref. [16], i.e., it corresponds to a highly agglomerated alumina powder of Ref. [38].

To identify the various simulated compaction processes, the values of the coefficients $\kappa_x$ and $\kappa_y$ are used, and the process is denoted by the pair "$\kappa_x;\kappa_y$". Computer experiments have been performed for the following processes:

1. «1;1» is the uniform compression. Strain increment $\Delta\varepsilon_{ij}$ and stress $\sigma_{ij}$ tensors are spherical, i.e., $\Delta\varepsilon_{ij} = (\varepsilon/3)\delta_{ij}$, $\sigma_{ij} = -p\,\delta_{ij}$, where $\varepsilon = \mathrm{Sp}(\Delta\varepsilon_{ij})$.
2. «0.9;1»: the intensity of the deviator $\gamma_{ij} = \Delta\varepsilon_{ij} - (\varepsilon/3)\delta_{ij}$ of the strain increment tensor is equal to $\gamma = |\varepsilon|\sqrt{6}/87$.
3. «0.75;1»: $\gamma = |\varepsilon|\sqrt{6}/33$.
4. «0.5;1»: $\gamma = |\varepsilon|\sqrt{6}/15$.
5. «0.25;1»: $\gamma = |\varepsilon|\sqrt{6}/9$.
6. «0;1» is the compression along the *Oy* and *Oz* axes: $\gamma = |\varepsilon|\sqrt{1/6}$.
7. «0;0.5»: $\gamma = |\varepsilon|\sqrt{2}/3$.
8. «0;0» is the uniaxial compression along the *Oz* axis: $\gamma = |\varepsilon|\sqrt{2/3}$.
9. «0;–0.1» is the compression on the *Oz* axis with simultaneous slight stretching on the *Oy* axis: $\gamma = |\varepsilon|\sqrt{74}/9$.
10. «0;–0.2»: $\gamma = |\varepsilon|\sqrt{31/24}$.
11. «0;–0.3»: $\gamma = |\varepsilon|\sqrt{278/147}$.

In addition, three processes of the form "$\kappa_n,\kappa_n$" have been modeled with values

$$\kappa_1 = \frac{1}{4}; \qquad \kappa_2 = \frac{\sqrt{3}-1}{\sqrt{3}+2} \cong 0.196; \qquad \kappa_3 = \frac{-1}{2}\frac{\sqrt{31}-4}{\sqrt{31}+2} \cong -0.103, \qquad (8)$$

which in terms of the ratio $\gamma/\varepsilon$ are analogs, respectively, of the processes "0;1", "0;0.5" and "0;–0.2".

### 3. Compaction curves

In all the processes listed in the previous section, the model cell has been compacted to a specified level $p_{max}$ = 5 GPa of external load along the *Oz* axis, i.e., until the condition $p_z = p_{\max}$. Then the model cell has been unloaded, during which the cell has been deformed in all directions with the same relative rates as when compacting, but the opposite sign. Figure 1 shows the compaction and unloading curves $p_z(\rho)$ of the five processes. Note that several (from 6 to 10) statistically independent calculations have been performed to construct each curve, after which the calculated data have been averaged. The stage of unloading, or "elastic" unloading [17, 19, 24], is characterized by a change $\Delta\rho_u$ in density. However, in addition to the purely elastic unloading interparticle contacts, pressure relief is also accompanied by irreversible processes of relative movement of particles. Therefore, the name of these stages "elastic" is quite conditional, and only assumes that elastic processes are likely to prevail here.

Figure 1 shows that at the loading stage, the "density – maximum pressure" compaction curves are very close to each other. The processes "1;1", "0.5;1", and "0;1" coincide within the calculation error (about 0.3%), and the deviation from them of the process "0;0"(uniaxial compression) is about 1% in density. Insensitivity of nanopowder compaction to the loading scheme was noted earlier in the works [16-18]. The reason for this insensitivity seems to be the mutual compensation of two opposite effects. On the one hand, the transition to asymmetric loading (from uniform compression "1;1" to biaxial "0;1" and then to uniaxial "0;0") leads to a given level of external (maximum) pressure ($p_z$) to reduce the average pressure in the powder, which should reduce the density of the compact. So, for a uniaxial process "0;0" at $p_z$ = 5 GPa, the calculated hydrostatic pressure is only $p$ = 3.9 GPa. On the other hand, an increase in shear deformations and stresses characterized by the deviators intensity of strain ($\gamma$) and stresses ($\tau$) contributes to the achievement of higher densities. If for uniform compression $\gamma, \tau \equiv 0$, then for a uniaxial process $\gamma = 0.816$, and the value of $\tau$ at maximum pressure, as shown by numerical estimates, reaches 1.3 GPa.

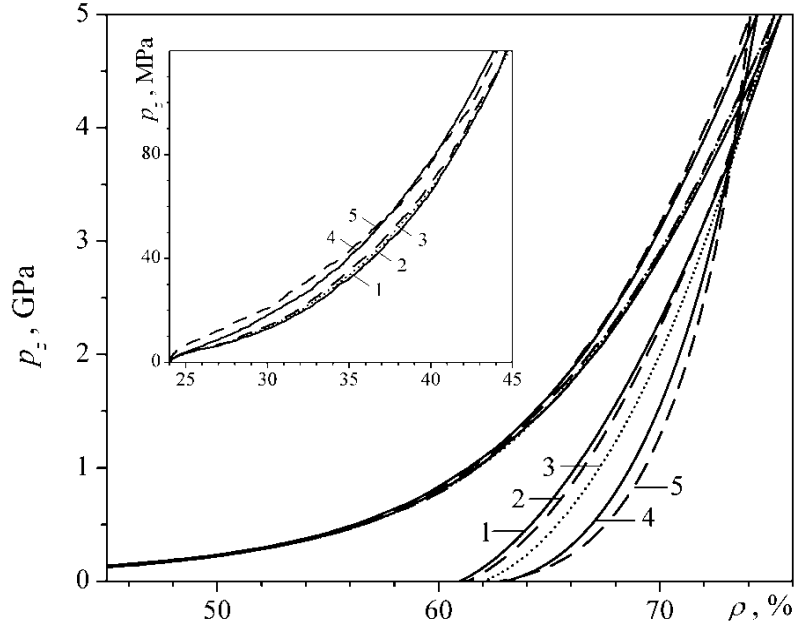

**Fig. 1.** Compaction curves in the "density–pressure $p_z$" plane for the processes «1;1» (solid line 1), «0.5;1» (dashed line 2), «0;1» (dotted line 3), «0;0» (solid line 4) and «0;–0.3» (dashed line 5). The insert shows the initial stage of the compaction at an enlarged scale.

At the stages of "elastic" unloading, the difference in density between the simulated processes becomes more noticeable, and is about 2% between the uniaxial and uniform processes when the external pressure is completely released. Moreover, if the uniaxial process is characterized by lower density values during loading, then after the "elastic" unloading, the lower density corresponds to the process of uniform compression-stretching. Density changes $\Delta \rho_u$ at the unloading stages for processes "1;1" and "0;0" are 14.5% and 11.3%, respectively.

Separately, the question of the significance of the third invariants of strain and stress tensors for the description of a powder body has been investigated. As noted [12], for a wide class of isotropic materials, for example, classical elastic and viscous bodies, scalar mechanical properties are determined only by the first two invariants of these tensors. This approximation fully justifies itself when describing powders of micron and larger sizes [12, 19, 20]. However, for nanoscale powders, this approximation has not yet been verified. In order to perform this check we have analyzed the processes "0;1", "0;0.5" and "0;–0.2" in comparison with their analogues, i.e., with processes "$\kappa_n, \kappa_n$" (8), see Fig. 2. Despite the equality of the relation $\gamma / \varepsilon$, the analyzed pairs of processes are characterized by different values of the third invariant of the strain tensor.

If we use the value $I_3 = \|\gamma_{ij}\|^{1/3} / \varepsilon$ as an additional characteristic (where $\|\gamma_{ij}\| = \gamma_{xx}\gamma_{yy}\gamma_{zz}$ is the determinant of the deformation deviator), then for the pair of processes "0;1" and "$\kappa_1,\kappa_1$" we have $I_3 = -0.21$ and $+0.21$, respectively; for the pair "0;0.5" and "$\kappa_2,\kappa_2$" – 0 and 0.24; for the pair "0;–0.2" and "$\kappa_3,\kappa_3$" – 0.84 and 0.58. As a consequence, these pairs are characterized by different values of individual components of the stress tensor and the similar third invariant of the stress deviator. So, for the pair "0;0.5" and "$\kappa_2,\kappa_2$" at a pressure $p_z = 5$ GPa, the

compact density reaches $\rho \cong 74.5\%$, and the "side" pressures in the process "0;0.5" are $p_x \cong 3.6$ GPA and $p_y \cong 4.3$ GPA, and in the process "κ2,κ2" $p_x \equiv p_y \cong 3.8$ GPa. Nevertheless, despite the obvious difference in the stress state realized in these processes, the curves of their compaction in the invariant variables $p(\rho)$ and $\tau(\rho)$ coincide within the calculation error. The dependences of the stress deviator intensity on the compact density of the analyzed processes are shown in Fig. 2. It can be seen that the curves $\tau(\rho)$ of each pair are satisfactorily consistent both at the loading and unloading stages. The curves $p(\rho)$ show a similar agreement. Thus, the coincidence of the curves $p(\rho)$ and $\tau(\rho)$ for processes with different values of the third invariants of strain and stress tensors confirms the hypothesis traditionally used in the plasticity theory that the first two invariants of these tensors are sufficient to describe nanoscale powder systems.

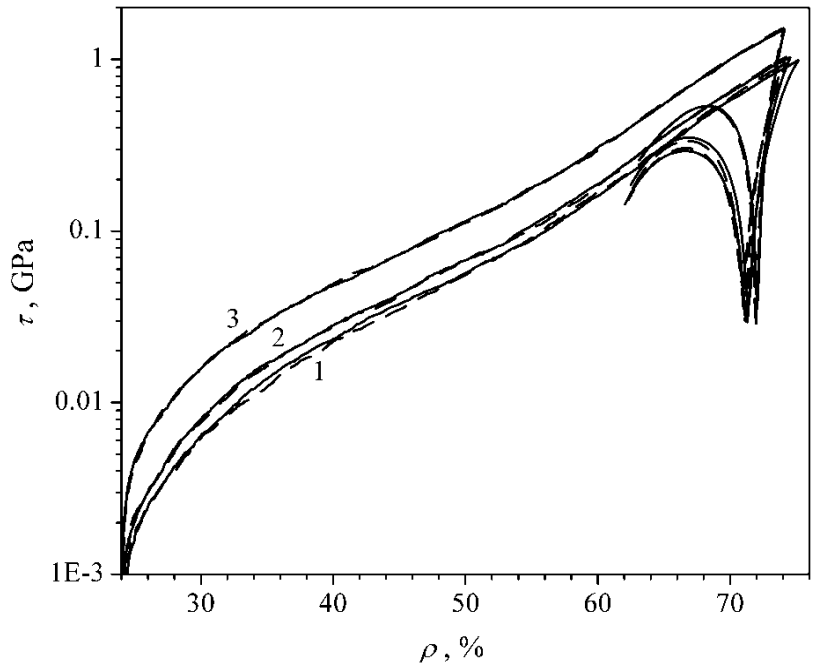

**Fig. 2.** The dependence of the stress deviator intensity on the compact density for processes: "0;1", "0;0.5" and "0;–0.2" (solid lines 1, 2 and 3, respectively); "κ1,κ1", "κ2,κ2" and "κ3,κ3" (dashed lines 1, 2 and 3, respectively).

**4. Extraction of "elastic-reversible" contribution**

In Refs. [17, 18] the elastic contribution to the total change in the compact density $\Delta\rho_e$ at the compaction stage was identified with the value of $\Delta\rho_u$, i.e., with the change in density recorded at the stage of external pressure relief. However, this identification is not strict, since at the stage of pressure relief in the powder system, simultaneously with the processes of elastic unloading of interparticle contacts, the processes of relative (tangential) slippage of particles inevitably occur that corresponds from the macroscopic point of view to the process of plastic flow of the material. For the same reason, the estimation of the uniform compression modulus

$$K_p = \rho \frac{dp}{d\rho}\bigg|_{ela}, \qquad (9)$$

made in Ref. [39] by the slope of the $p_z(\rho)$ curve at the initial pressure relief section can only be considered a "bottom estimate". It is also worth noting the high labor cost of such an as-sessment of elastic properties. Analysis of their changes with increasing compact density requires the simulations of a large number of elastic unloading branches [17, 39].

In connection with what has been said in the present work, a fundamentally different way of evaluating the elastic properties of a compact has been realized. The elastic stress increment $\Delta \sigma_{ij}$ in the simulated system, corresponding to a small density increment $\Delta \rho$, was "measured" at each step of deformation of the model cell immediately after changing its size and proportional increment of the corresponding coordinates of all particles. Only after this measurement did the relaxation mechanism "turn on", i.e. particles began to move to new equilibrium positions. Thus, the processes of mutual slippage are separated from purely elastic deformation. It should be noted that the relative displacements of particles proportional to the deformation of the model cell correspond to the well-known Voigt approximation for the strain field in a continuous medium [28].

Taking as an assumption the powder compact isotropy, we can use the Hooke law in the form [40, 41]

$$\Delta \sigma_{ij} = K_p \varepsilon + 2\mu_e \left( \Delta \varepsilon_{ij} - \frac{\varepsilon}{3} \delta_{ij} \right), \qquad 2\mu_e = 3K_p \frac{1-2\nu_p}{1+\nu_p}, \tag{10}$$

to describe its elastic properties that allows us to determine the elastic moduli $K_p$ and $\nu_p$ of the powder body. The expression for the compression module is written above, see Eq. (9), and the Poisson's ratio $\nu_p$ for the simulated processes can be determined by the ratio of the increments of the stress tensor components, for example,

$$\text{«0.5;1»:} \left. \frac{\Delta p_x}{\Delta p_z} \right|_{ela} = \frac{1+3\nu_p}{2+\nu_p} ; \quad \text{«0;1»:} \left. \frac{\Delta p_x}{\Delta p_z} \right|_{ela} = 2\nu_p ; \quad \text{«0;0»:} \left. \frac{\Delta p_x}{\Delta p_z} \right|_{ela} = \frac{\nu_p}{1-\nu_p}. \tag{11-a}$$

Note that for the "1;1" process (uniform compression) the Poisson's ratio cannot be determined, and for processes with three different diagonal components of the strain tensor, for example, the "0;0.5" process, the Poisson's ratio can be determined using different pairs of stress tensor components:

$$\text{«0;0.5»:} \left. \frac{\Delta p_x}{\Delta p_z} \right|_{ela} = \frac{3\nu_{p,x}}{2-\nu_{p,x}} ; \qquad \left. \frac{\Delta p_y}{\Delta p_z} \right|_{ela} = \frac{1+\nu_{p,y}}{2-\nu_{p,y}}. \tag{11-b}$$

In the case of validity of the isotropy approximation, the values $\nu_{p,x}$ and $\nu_{p,y}$ must coincide. Figures 3 and 4 show the elastic modules obtained using expressions (9) – (11).

Figure 4 shows that the Poisson's ratio cannot be considered as an unambiguous function of the compact density, since we observe different dependences $\nu_p(\rho)$ for different processes, as well as the mismatch of values $\nu_{p,x}$ and $\nu_{p,y}$ for the process "0;0.5". This indicates that the elastic properties of the simulated system cannot be described by law (10) with two elastic modules, i.e., the approximation of the material isotropy for the compacted powder is not performed. The calculated data show that the distribution of the directions of interparticle contacts in the powder system is isotropic within the calculation error in all simulated processes, but the distribution of contact forces has a noticeable angular dependence (see, for example, Fig. 14 in Ref. [16]). The latter, together with the nonlinear law of elastic particle interaction (4), apparently leads to a significant anisotropy, induced by the external loading conditions, of the elastic properties of the compact.

Despite the induced anisotropy noted above, the compression modulus $K_p$, determined by the first invariants of the strain and stress tensors and presented in Fig. 3, is determined for most simulated processes quite unambiguously, and is satisfactorily approximated by the common expression

$$K(\rho) = k_1\rho + k_2\rho \exp(k_3\rho), \qquad (12)$$

with coefficients $k_1$ = 10.987 GPa, $k_2$ = 0.638 GPa, and $k_3$ = 5.744. The only exceptions are processes with extension along the $Oy$ axis. Here, apparently, the proximity to the fracture surface of the powder body manifests itself. This surface according to studies [17, 25] is located on the "$p$–$\tau$" plane somewhat to the left of the $\tau(p)$ curve of the uniaxial compression process "0;0".

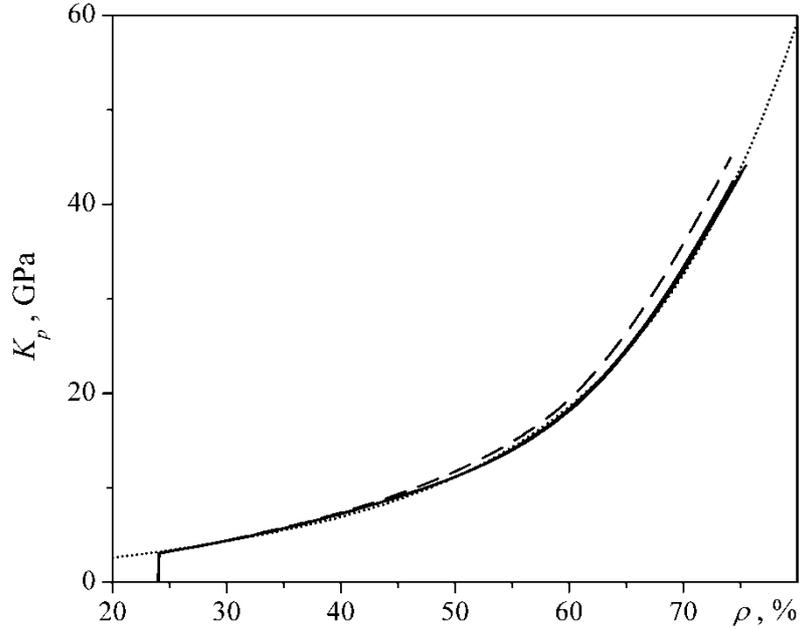

**Fig. 3.** Compression module, calculated according Eq. (9), for the processes "1;1", "0;1", "0;0" (solid lines from bottom to top, almost indistinguishable) and "0; –0.3" (dashed line). Dotted line is the approximation (12).

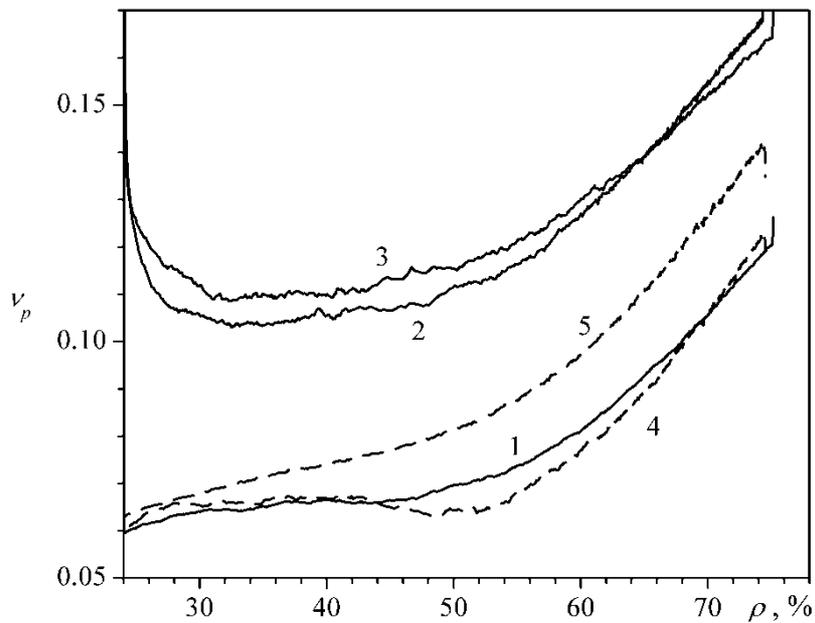

**Fig. 4.** Poisson's ratio calculated by Eq. (11), for the processes "0.5;1", "0;1" and "0;0" (solid lines 1, 2, and 3, respectively), as well as the coefficients $v_{p,x}$ (dashed line 4) and $v_{p,y}$ (dashed line 5) for process "0; 0.5".

With a known dependence $K_p(\rho)$, the elastic component of the density increment $\Delta\rho_{ela}$ accumulated during loading can be calculated using the equation:

$$\Delta\rho_{ela} = \int_0^p \frac{\rho}{K_p(\rho)} dp. \qquad (13)$$

Numerically calculating the recorded integral along the loading curve $p(\rho)$, we obtain the relationship of $\Delta\rho_{ela}$ with the parameters that determine the state of the powder compact ($\rho$, $p$, $p_{max}$, etc.). Figure 5 shows the dependencies $\Delta\rho_{ela}(p)$ thus obtained. The noticeable difference in the $\Delta\rho_{ela}(p)$ curves for processes with almost identical dependencies $K_p(\rho)$ is due to the difference in the compaction curves $p(\rho)$, which are implicitly included in the subintegral expression of Eq. (13).

The $\Delta\rho_{ela}(p_{max})$ curves have the same non-linear character as the curves in Fig. 5, and are qualitatively consistent with the corresponding dependencies presented in Ref. [17]. In quantitative terms, the obtained values of elastic deformation $\Delta\rho_{ela}$ are below the values for the model system II with similar parameters in Ref. [17], where the elastic part of the density increment has been identified with the value of $\Delta\rho_u$. Thus, at a pressure of $p_{max} = 5$ GPa for the process of uniform compression of system II $\Delta\rho_{ela} = 14.65\%$ is obtained in Ref. [17], while for the process "1;1" we have $\Delta\rho_{ela} = 12.68\%$. The decrease in the calculated values of elastic deformation is due to the exclusion of the contribution of plastic processes (mutual slippage of particles) in the algorithm for calculating elastic properties used in this work.

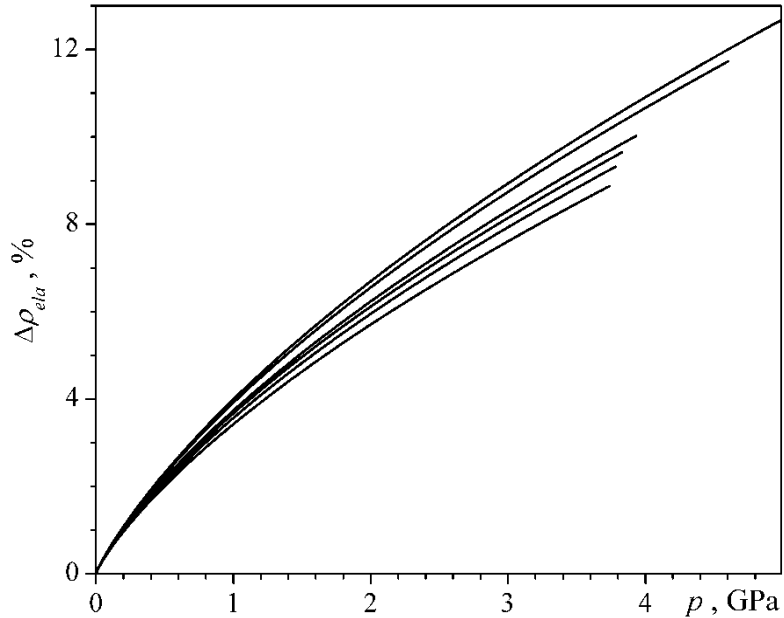

**Fig. 5.** The elastic part of the compact density increment calculated from the calculated compaction curves and Eqs. (9) and (13), depending on the hydrostatic pressure $p$ for the processes (lines from top to bottom): «1;1», «0;1», «0;0», «0;–0.1», «0;–0.2», and «0;–0.3».

The obtained dependences $\Delta\rho_{ela}(p)$ allow us to distinguish an irreversible (plastic) component $\rho_{plast} = \rho - \Delta\rho_{ela}$ from the total deformation $\rho(p)$ of the model cell. The dependence of the $\rho_{plast}$ value on the external pressure is shown in Fig. 6. There, for comparison, the original compaction curves $\rho(p_{max})$ containing the elastic contribution $\Delta\rho_{ela}$ are presented. The dependencies $\rho_{plast}(p_{max})$, as well as the original dependencies corresponding to different processes, are quite close to each other. It is interesting to note that at relatively low pressures ($p_{max} \leq 100$ MPa) the highest values of the achieved density (both $\rho$ and $\rho_{plast}$) are realized in the process of uniform compression "1;1", but in the range of high pressures ($p_{max} > 1$ GPa), due to the large values of the elastic contribution $\Delta\rho_{ela}$, this process is characterized by the lowest values of the plastic component $\rho_{plast}$. The difference in $\rho_{plast}$ values between uniaxial and uniform compression processes reaches 1.5% at $p_{max} = 5$ GPa. In the hypothetical limit of infinitely high pressures ($p_{max} \to \infty$), the $\rho_{plast}(p_{max})$ dependencies are close to the power form $\rho_{plast} = \rho_\infty - k_\rho / p_{max}^{1/2}$, which allows us to estimate the maximum possible density $\rho_\infty$ of compacts. For the studied processes, as shown in the inset of Fig. 6, it lies in the range from 67% (uniform compression "1;1") to 72% (process "0;–0.3").

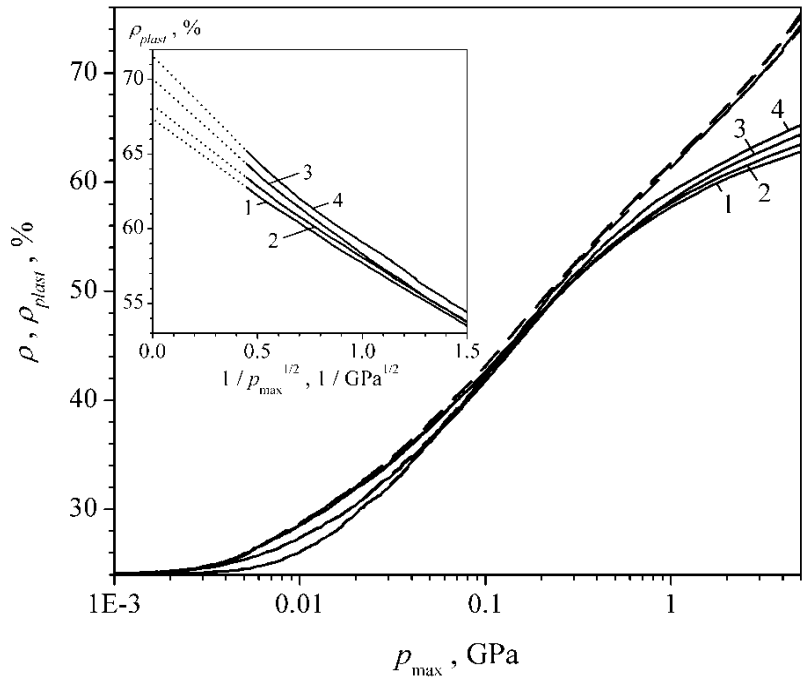

**Fig. 6.** The compact density $\rho$ (dashed lines) and the plastic irreversible contribution $\rho_{plast}$ (solid lines) depending on the maximum external pressure ($p_z$, along the $Oz$ axis) for the processes "1;1", "0;1", "0;0", and "0;–0.3" (lines 1, 2, 3, and 4, respectively). Inset: $\rho_{plast}$ values at high pressures and approximation to the limit $p_{max} \to \infty$ (dashed lines).

## 5. The yielding surface
The key parameter of the deformable body when describing its mechanical properties in the framework of plasticity theories [12-15, 19, 20] is the yielding surface, which in the space of the stress tensor component determines the boundary between the elastic deformation region and the plastic flow region. In contrast to plastically incompressible materials, in particular, compact metals, the powder yielding surface must depend not only on the intensity of the stress deviator ($\tau$), but also on the value of the first invariant of the stress tensor ($p$), and also on the current density as a parameter, passing in the limit of the pore-free state in the flow condition of the solid material. As the density, in accordance with the accepted analogy, it is necessary to understand the value of $\rho_{plast}$, i.e., the accumulated "plastic" deformations without the elastic contribution.

Many researchers use an approximation of the yielding surface equation in the form of an ellipse in the coordinates "$p - \tau$" to describe the behavior of porous bodies whose reaction to a change in the load sign can be neglected [12-15, 19]. For nanopowder compacts, as noted in previous works [17, 18], the elliptical surface is not applicable. If the sintered porous body is characterized by the presence of formed contacts between the particles, which is why it is able to resist almost equally tensile and compressive deformations, then the powder has a relatively weak resistance to tensile deformations. The consequence of this is a significant distortion of the flow ellipse and a noticeable shift towards positive values of hydrostatic pressure [17, 18].

Another feature of the powder body noted in Refs. [17, 25] is the presence of a fracture surface in the region of processes with tensile stresses. In the space of invariants "$p - \tau$", the fracture surface is located slightly to the left of the curve $\tau(p)$, corresponding to the uniaxial compression "0;0". In this regard, in the present work we used processes with only a relatively small stretch ("0;–0.1", "0;–0.2", and "0;–0.3"), located on the plane "$p - \tau$" near the uniaxial compression curve. In this case, the area of compressive deformations, ranging from uniaxial to uniform compaction, which is responsible for the so-called consolidation locus of yield surface [19], is mainly analyzed.

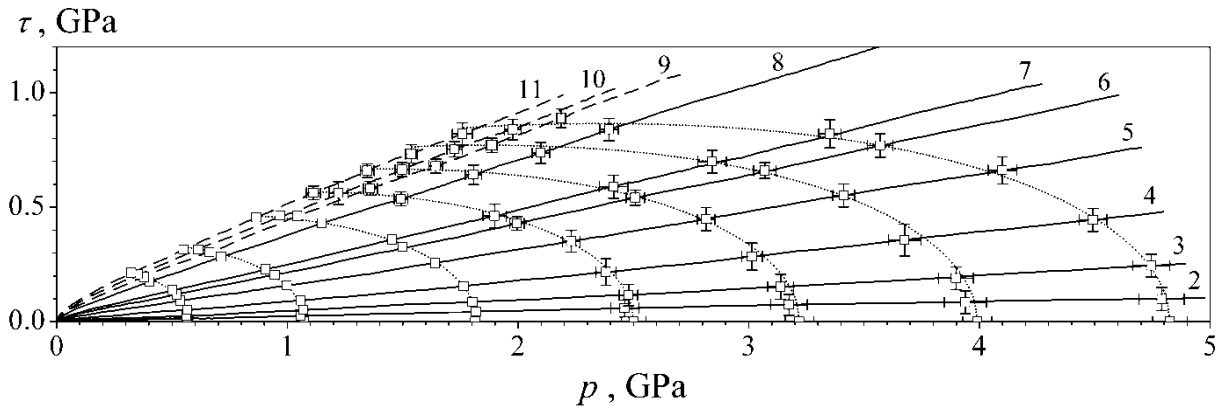

**Fig. 7.** Dependences of stress deviator intensity on hydrostatic pressure for simulated processes (line numbers correspond to process numbers in section 2). Dots mark the states corresponding to the values $\rho_{plast}$ = 0.55, 0.58, 0.60, 0.61, 0.617, 0.622, and 0.627. The dashed lines indicate the position of the yielding surface by Eq. (14) for these values.

Figure 7 shows the compression curves for the simulated processes in the space of invariants "$p - \tau$", as well as the points ($p,\tau$) on these curves, corresponding to the given densities $\rho_{plast}$. For 4 larger $\rho_{plast}$ values (0.61, 0.617, 0.622, and 0.627) the errors of statistical averaging of the calculated data are also shown. At lower densities, the errors become comparable, or even less, of the size of the symbols in the figure. The uniform compression curve "1;1" (line 1) in these coordinates is located along the abscissa. Within the statistical error, all the calculation points presented in Fig. 7 are satisfactorily approximated by the common dependence $\tau(p)$ for the yielding surface in the form:

$$\frac{\tau}{p_a} = \left[\tau_1 + \tau_2\left(\frac{p}{p_a}\right)\right]\sqrt{1 - \frac{p}{p_a}} \qquad (14)$$

where $p_a$ is the value of $p$ corresponding to the density $\rho_{plast}$ at uniform compression (the process «1;1»), i.e., the coordinate of the intersection of the yielding surface with the hydrostatic pressure axis; and the coefficients $\tau_1$ and $\tau_2$ depend on the density as follows: $\tau_1 = 2.40 - 3.62\rho_{plast}$, $\tau_2 = 0.51 - 0.42\rho_{plast}$. The iso-lines determined by Eq. (14) are shown in Fig. 7 by dotted lines. It can be seen that the location and shape of the levels of the yielding surface corresponding to given densities $\rho_{plast}$ generally confirms the convexity and smoothness (without corner points) of the powder compact yielding surface in the form of a shifted and deformed ellipse.

## 6. Flow rule of oxide nanopowders

The calculated compaction curves and the constructed family of iso-lines ($\rho_{plast}$ = const) of the yielding surface, shown in Fig. 7, allow a detailed analysis of the powder body flow. In the framework of the plastic body phenomenology, the hypothesis of the "associated rule" is widely used [12-14, 19, 20], according to which the strain increment should be normal to the yielding surface in the space of the stress tensor components. Moreover, under the strains it is necessary to understand precisely the plastic parts of the total strains.

The increment of the total strains $\Delta\varepsilon_{ij}$ in the simulated processes is determined by the deformation of the model cell. Separating them into elastic and plastic parts, i.e., $\Delta\varepsilon_{ij} = \Delta\varepsilon_{ij}^{(e)} + \Delta\varepsilon_{ij}^{(p)}$, we define the elastic parts by the relations:

$$\Delta\varepsilon_{xx}^{(e)} = -\frac{\Delta p_x}{3K_p}, \qquad \Delta\varepsilon_{yy}^{(e)} = -\frac{\Delta p_y}{3K_p}, \qquad \Delta\varepsilon_{zz}^{(e)} = -\frac{\Delta p_z}{3K_p}, \qquad (15)$$

which, as can be easily seen, correspond to expression (9) for the elastic modulus of compression. Subtracting the elastic parts determined by Eqs. (15) from the total strains, we obtain plastic strains $\Delta\varepsilon_{ij}^{(p)}$ that can be used to verify the phenomenological theory of a plastic body, and in particular, the associated rule.

One of the consequences of the associated rule is the coaxiality of deviators of stress tensor and strain increments tensor, i.e., $\tau_{ij} \propto \gamma_{ij}^{(p)}$. In the case of triaxial loading with given values of the diagonal components of the strain increment tensor, the coaxiality gives a relationship between the diagonal values of the stress tensor, which can be used to determine, for example, $p_y$ component from the known values of $p_x$ and $p_z$:

$$p_y^{(ass)} = \frac{\Delta\varepsilon_{zz}^{(p)} - \Delta\varepsilon_{yy}^{(p)}}{\Delta\varepsilon_{zz}^{(p)} - \Delta\varepsilon_{xx}^{(p)}} p_x + \frac{\Delta\varepsilon_{yy}^{(p)} - \Delta\varepsilon_{xx}^{(p)}}{\Delta\varepsilon_{zz}^{(p)} - \Delta\varepsilon_{xx}^{(p)}} p_z. \qquad (16)$$

As can be seen from the presented relation, when two diagonal components of the strain tensor are equal, the corresponding diagonal components of the stress tensor also coincide. Thus, for most of the simulated processes, the coaxiality is guaranteed by the symmetry conditions: equal stresses in the directions with equal strain rates.

All three diagonal components of the strain tensor differ in the processes "0;0.5", "0;–0.1", "0;–0.2", and "0;–0.3". A comparison of the calculated pressure $p_y$ and the values $p_y^{(ass)}$ determined by Eq. (16) is presented for these processes in Fig. 8. The figure shows a fairly good agreement between the quantities $p_y$ and $p_y^{(ass)}$, i.e., within the simulation error, it is possible to state the fulfillment of the coaxiality of the deviators of strains and stresses.

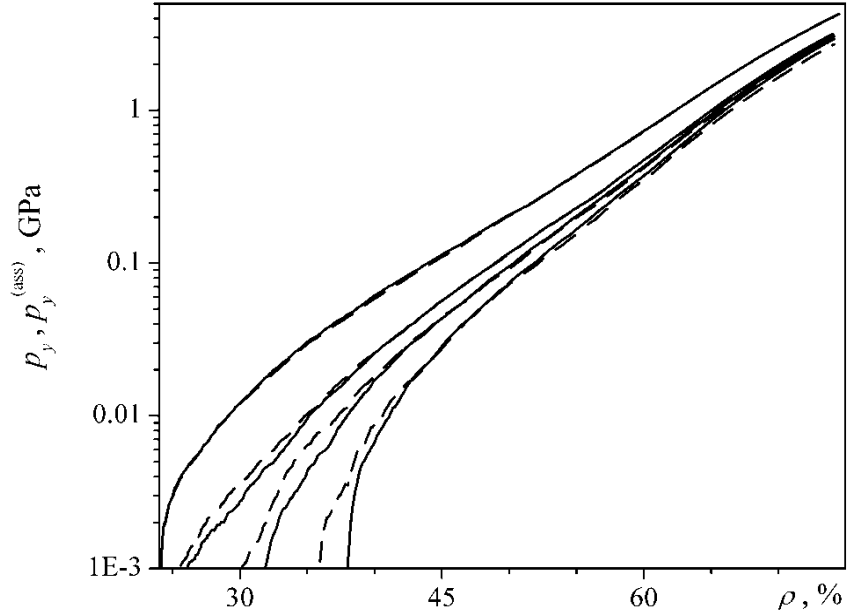

**Fig. 8.** The dependence of the pressure along the $Oy$ axis on the compact density for the processes (from top to bottom) "0;0.5", "0;–0.1", "0;–0.2", "0;–0.3". Solid lines are the simulation curves, dashed lines are the curves $p_y^{(ass)}$ calculated by Eq. (16).

As applied to the invariants of strain and stress tensors, another interesting consequence of the associated law for powder bodies is relation

$$(\varepsilon^{(p)}, \gamma^{(p)}) = \lambda_1' \nabla \Phi , \qquad (17)$$

where the potential $\Phi$ is either the dissipative potential of the deformable body or its yielding function, whose iso-levels coincide with the levels of the yielding surface (14) presented in Fig. 7. As has been shown in Ref. [18], the choice of the yielding function as potential $\Phi$ is preferable, since the latter is a more visual, unambiguous, and reliable characteristic of a powder body. In particular, it has been established that the yielding surface is practically independent of the intermediate unloadings during compaction and, as a consequence, of the initial state determined by the initial density $\rho_0$ of the pressed powder.

Relation (17) requires (see Ref. [14]) that the direction of the plastic strain vector $(\varepsilon^{(p)}, \gamma^{(p)})$ be normal to the iso-levels of the yielding surface shown in Fig. 7. Figure 9 shows the directions of the vectors $(\varepsilon^{(p)}, \gamma^{(p)})$ and vectors $\nabla \Phi$ for two iso-lines, with $\rho_{plast} = 0.60$ and 0.627. We see that the fulfillment of the associated rule (17), i.e., the collinearity of the vectors $(\varepsilon^{(p)}, \gamma^{(p)})$ and $\nabla \Phi$ is observed only in the trivial case of uniform compression. The slightest deviation from uniform conditions, already for the "0.9;1" process, demonstrates a noticeable violation of collinearity (17). In this case, the strain vector $(\varepsilon^{(p)}, \gamma^{(p)})$ for all processes deviates from $\nabla \Phi$ towards the compaction curve $\tau(p)$, the direction of which can be determined by vector $(\Delta p, \Delta \tau)$.

The noted feature allows us to formulate an alternative criterion for the flow of oxide nanopowders in the form [18]:

$$(\varepsilon^{(p)}, \gamma^{(p)}) = (1-\omega) \lambda_1' \nabla \Phi + \omega \lambda_2 (\Delta p, \Delta \tau) , \qquad (18)$$

where $\omega$ is the weight coefficient determining the effect of the ongoing process on the "direction" of strains initiated in the powder system; $\lambda_2$ is the dimensional coefficient defined as follows,

$$\lambda_2 = \lambda_1' \frac{|\nabla\Phi|}{|(\Delta p, \Delta \tau)|} = \lambda_1' \sqrt{\left[\left(\frac{\partial \Phi}{\partial p}\right)_\tau^2 + \left(\frac{\partial \Phi}{\partial \tau}\right)_p^2\right] / \left[\Delta p^2 + \Delta \tau^2\right]}. \tag{19}$$

Passing from the relation for invariants (18) to the strain and stress tensors, we obtain the general form for writing the flow rule of a nanopowder body:

$$\Delta \varepsilon_{\alpha\beta} = (1-\omega)\lambda_1' \cdot \frac{\partial \Phi}{\partial p^{\alpha\beta}} + \omega \lambda_2 \left(\Delta p \frac{\partial p}{\partial p^{\alpha\beta}} + \Delta \tau \frac{\partial \tau}{\partial p^{\alpha\beta}}\right), \tag{20}$$

where

$$\frac{\partial \Phi}{\partial p^{\alpha\beta}} = \left(\frac{\partial \Phi}{\partial p}\right)_\tau \frac{\partial p}{\partial p^{\alpha\beta}} + \left(\frac{\partial \Phi}{\partial \tau}\right)_p \frac{\partial \tau}{\partial p^{\alpha\beta}}, \qquad \frac{\partial p}{\partial p^{\alpha\beta}} = \frac{\delta_{\alpha\beta}}{3}, \qquad \frac{\partial \tau}{\partial p^{\alpha\beta}} = \frac{1}{\tau}\left(p_{\alpha\beta} - p \cdot \delta_{\alpha\beta}\right),$$

$\delta_{\alpha\beta}$ is the symbol of Kronecker. The first term on the right in relation (20) corresponds to the associated rule, and the second term determines the influence of the process. Figure 9 shows that the weight coefficient of this influence $\omega$ is not constant. With approaching the uniform process, the strain vector $(\varepsilon^{(p)}, \gamma^{(p)})$ becomes much closer to the vector $(\Delta p, \Delta \tau)$, i.e., the weight coefficient becomes close to unity.

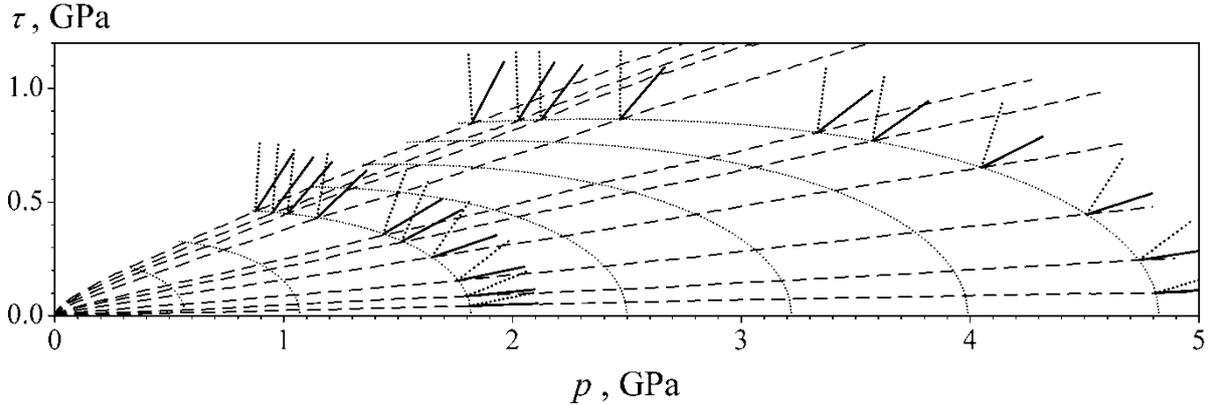

**Fig. 9.** Compaction curves and contours of the yielding surface (the lines are the same as in Fig. 7). The arrows at the intersections of the compaction and contour curves for $\rho_{plast} = 0.60$ and 0.627 indicate the direction of the vector $\nabla\Phi$, i.e., the normal to the yielding surface (upper dashed arrows), and the direction of the vector $(\varepsilon^{(p)}, \gamma^{(p)})$, which determines the deformation, (lower solid arrows).

It can be assumed that the quantity $\omega$ is a function of the derivative $\tau_p = d\tau/dp$ along the curve $\tau(p)$, which determines the compaction process. Introducing unit vectors

$$v = \frac{(e,\gamma)}{|(e,\gamma)|}, \qquad n = \frac{\nabla\Phi}{|\nabla\Phi|}, \qquad t = \frac{(1,\tau_p)}{|(1,\tau_p)|},$$

and requiring vanishing of the vector product of vectors $v$ and $(1-\omega)n + \omega t$, for calculating $\omega$ we obtain the relation:

$$\omega = \frac{v_1 n_2 - v_2 n_1}{v_1(n_2 - t_2) - v_2(n_1 - t_1)}. \tag{21}$$

The values of the coefficient $\omega$ as a function on the ratio $\tau_p$ calculated according to this relation for the intersections of the compression curves of all the studied processes with 7 iso-lines shown in Fig. 7 and 9 are presented in Fig. 10. Analysis of the obtained data shows that the function does not depend on $p_{plast}$, i.e., is common to all iso-lines, and is satisfactorily approximated by the expression:

$$\omega = \omega_1 - \omega_2 \tau_p, \tag{22}$$

with coefficients $\omega_1 = 0.9$ and $\omega_2 = 1.0$. The Eq. (22) closes the system of previous relations that determine the change in the components of the strain tensor of the powder compact under a given external action, i.e., for a given increment of the stress tensor components.

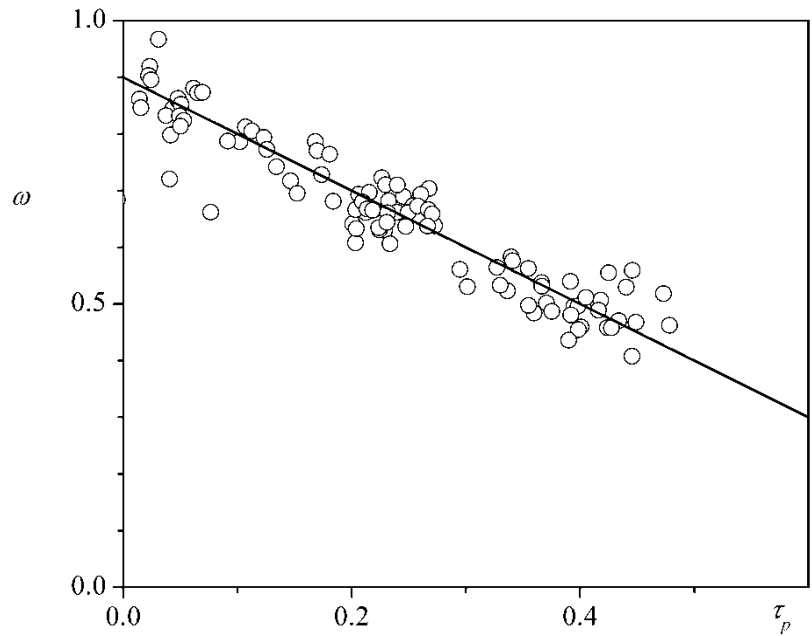

**Fig. 10.** The parameter $\omega$, which determines in accordance with Eqs. (18)–(20) the direction of the plastic strain vector, depending on the ratio of the rate of change of the stress tensor invariants.

## 7. Conclusion

For a model system that corresponds to a nanosized powder of aluminum oxide, compaction curves have been calculated for compaction under various conditions. The processes of uniform compression, biaxial and uniaxial pressing, as well as a number of processes of trilateral non-uniform compression, i.e., with different compression speeds along different directions, have been studied. The elastic properties of powder compacts are investigated. The invariance of the compression modulus and the non-constancy of the Poisson's ratio are revealed, which indicates the inapplicability of the approximation of an isotropic body to a powder compact and, as a consequence, the insufficiency of two elastic modules to describe its elastic properties. Nevertheless, the reliably set value of the compression modulus allows one to determine the elastic and plastic parts of the density increment and the strain tensor.

The theory of a plastically compacted porous body [12–15], traditionally being used to describe powder compacts, has been verified. The insensitivity of the mechanical properties of the powder body on the third invariants of the strain and stress tensors, as well as the coaxiality of the strain and strain deviators are established that is fully consistent with traditional theoretical concepts. However, in general, it should be recognized that the well-known associ-ated rule of plastic flow does not apply to nanosized oxide powders, since in the space of stress tensor invariants the vector of invariants of the plastic strain increment tensor turns out to be non-normal to the iso-levels of the yielding surface. Instead of the traditional associated rule, another criterion is proposed that allows us to predict the nature of deformation processes in the system. According to the proposed criterion, the form of the strain increment tensor is determined not only by the direction of the gradient vector of the yielding function (associated rule), but also by the direction of the "vector", which determines the change in the components of the stress tensor during the compaction process. The ratio of contributions from these two vectors is determined by the weight coefficient $\omega$. A complete system of relations is formulated that uniquely determines the change in the components of the strain tensor in a system for a given external action. In addition to the flow criterion, this system contains an approximation of the iso-levels of the yielding surface and the dependence of the weight coefficient $\omega$ on the invariants of the stress tensor.

*The work was carried out with partial financial support of the RFBR (project 16-08-00277).*

**References**
[1] J.R. Pickens // J. Mater. Sci. **16** (1981) 1437.
[2] M. Bengisu, O.T. Inal // J. Mater. Sci. **29** (1994) 4824.
[3] C. Pecharroman, G. Mata-Osoro, L.A. Diaz, R. Torrecillas, J.S. Moya // Opt. Express. **17** (2009) 6899.
[4] N. Roussel, L. Lallemant, J.Y. Chane-Ching, S. Guillemet-Fristch, B. Durand, V. Garnier, G. Bonnefont, G. Fantozzi, L. Bonneau, S. Trombert, D. Garcia-Gutierrez // J. Am. Ceram. Soc. **96** (2013) 1039.
[5] L. Shen, C. Hu, S. Zhou, A. Mukherjee, Q. Huang // Opt. Mater. **35** (2013) 1268.
[6] E.H. Penilla, Y. Kodera, J.E. Garay // Adv. Funct. Mater. **23** (2013) 6036.
[7] J. Luo, J. Xu, In: *Proceedings of the 5th International Conference on Advanced Design and Manufacturing Engineering (ICADME 2015)* (Atlantis Press, Shenzhen, 2015), p. 99.
[8] C. Zhou, B. Jiang, J. Fan, X. Mao, L. Pan, Y. Jiang, L. Zhang, Y. Fang // Ceram. Int. **42** (2016) 1648.
[9] A.S. Kaygorodov, V.V. Ivanov, V.R. Khrustov, Yu.A. Kotov, A.I. Medvedev, V.V. Osipov, M.G. Ivanov, A.N. Orlov, A.M. Murzakaev // J. Eur. Ceram. Soc. **27** (2007) 1165.
[10] S.N. Bagayev, V.V. Osipov, V.A. Shitov, E.V. Pestryakov, V.S. Kijko, R.N. Maksimov, K.E. Lukyashin, A.N. Orlov, K.V. Polyakov, V.V. Petrov // J. Eur. Ceram. Soc. **32** (2012) 4257.
[11] V.P. Filonenko, L.G. Khvostantsev, R.Kh. Bagramov, L.I. Trusov, V.I. Novikov // Powder Metallurgy and Metal Ceramics. **31** (1992) 296.
[12] M.B. Shtern, G.G. Serdyuk, L.A. Maksimenko, Yu.V. Truhan, Yu.M. Shulyakov, *Phenomenological Theories of Powder Pressing* (Naukova Dumka, Kiev, 1982).
[13] E.A. Olevskii, M.B. Shtern // Powder Metallurgy and Metal Ceramics. **43** (2004) 355.
[14] A.L. Maximenko, E.A. Olevsky, M.B. Shtern // Comput. Mater. Sci. **43** (2008) 704.
[15] G.Sh. Boltachev, K.A. Nagayev, S.N. Paranin, A.V. Spirin, N.B. Volkov, In: *Nanomaterials: Properties, Preparation and Processes*, ed. by V. Cabral and R. Silva (Nova Science Publishers, Inc., New York, 2010), p. 1.
[16] G.Sh. Boltachev, K.E. Lukyashin, V.A. Shitov, N.B. Volkov // Phys. Rev. E **88** (2013) 012209.
[17] G.Sh. Boltachev, N.B. Volkov, E.A. Kochurin, A.L. Maximenko, M.B. Shtern, E.G. Kirkova // Granul. Mater. **17** (2015) 345.

[18] G.Sh. Boltachev, K.E. Lukyashin, A.L. Maximenko, R.N. Maksimov, V.A. Shitov, M.B. Shtern // Optical materials. **71** (2017) 145.
[19] J. Schwedes // Powder Technology. **11** (1975) 59.
[20] P.R. Nott // Acta Mechanica. **205** (2009) 151.
[21] I. Agnolin, J.-N. Roux // Phys. Rev. E **76** (2007) 061302.
[22] F.A. Gilabert, J.-N. Roux, A. Castellanos // Phys. Rev. E **75** (2007) 011303.
[23] F.A. Gilabert, J.-N. Roux, A. Castellanos A. // Phys. Rev. E **78** (2008) 031305.
[24] C. Salot, P. Gotteland, P. Villard // Granular Matter. **11** (2009) 221.
[25] P. Pizette, C.L. Martin, G. Delette, P. Sornay, F. Sans // Powder Technology. **198** (2010) 240.
[26] Yu.A. Kotov // J. Nanopart. Res. **5** (2003) 539.
[27] Yu.A. Kotov, V.V. Osipov, M.G. Ivanov, O.M. Samatov, V.V. Platonov, E.I. Azarkevich, A.M. Murzakaev, A.I. Medvedev // Technical Physics. **47** (2002) 1420.
[28] F. Nicot, F. Darve // Mechanics of Materials. **37** (2005) 980.
[29] H.C. Hamaker // Physica. **4** (1937) 1058.
[30] G.Sh. Boltachev, N.B. Volkov // Technical Physics. **56** (2011) 919.
[31] G.Sh. Boltachev, N.B. Volkov, N.M. Zubarev // Int. J. Solids Struct. **49** (2012) 2107.
[32] C. Cattaneo // Accademia dei Lincei, Rendiconti, Series 6. **27** (1938) 342.
[33] R.D. Mindlin // J. Appl. Mech. (Trans ASME) **16** (1949) 259.
[34] R.D. Mindlin, H. Deresiewicz // J. Appl. Mech. (Trans ASME) **20** (1953) 327.
[35] E. Reissner, H.F. Sagoci // J. Appl. Phys. **15** (1944) 652.
[36] J. Jäger // Archive of Applied Mechanics. **65** (1995) 478.
[37] A.I. Lur'e, *Three-Dimensional Problems of the Theory of Elasticity* (Interscience Publishers, NY, 1964).
[38] G.Sh. Boltachev, N.B. Volkov, A.S. Kaygorodov, V.P. Loznukho // Nanotechnologies in Russia. **6** (2011) 639.
[39] G.Sh. Boltachev, N.B. Volkov, E.S. Dvilis, O.L. Khasanov // Technical Physics. **60** (2015) 252.
[40] L.D. Landau, E.M. Lifshitz, *Theory of Elasticity* (Pergamon Press, Oxford, 1993).
[41] L.I. Sedov, *Mechanics of Continuous Media, vol. 2* (World Scientific, Singapore, 1997).